\documentclass[aps,prl,twocolumn,floatfix,superscriptaddress]{revtex4-1}
\usepackage{graphicx,txfonts,bm}
\usepackage[colorlinks,citecolor=magenta,linkcolor=blue]{hyperref}

\begin{document}

%% title
%%%%%%%%%%%%%%%%%%%%%%%%%%%%%%%%%%%%%%%%%%%%%%%%%%%%%%%%%%%%%%%%%%%%%%%%%%

\title{Pressure-Driven 5$f$ Localized-Itinerant Transition and Valence Fluctuation in Californium}

%% author list
%%%%%%%%%%%%%%%%%%%%%%%%%%%%%%%%%%%%%%%%%%%%%%%%%%%%%%%%%%%%%%%%%%%%%%%%%%

\author{Li Huang}
\email{lihuang.dmft@gmail.com}
\affiliation{Science and Technology on Surface Physics and Chemistry Laboratory, P.O. Box 9-35, Jiangyou 621908, China}

\author{Haiyan Lu}
\affiliation{Science and Technology on Surface Physics and Chemistry Laboratory, P.O. Box 9-35, Jiangyou 621908, China}

\date{\today}

%% abstract
%%%%%%%%%%%%%%%%%%%%%%%%%%%%%%%%%%%%%%%%%%%%%%%%%%%%%%%%%%%%%%%%%%%%%%%%%%

\begin{abstract}
A combination of the density functional theory and the single-site dynamical mean-field theory is employed to study the pressure dependence of electronic structure for cubic phase californium. We predict that its 5$f$ electrons could undergo an orbital-selective localized-itinerant transition under moderate pressure. The volume contraction causes remarkable charge redistribution and valence fluctuation behaviors, which are the driving forces of the divalent-trivalent transition. Additionally, we find that the angular momentum coupling mechanism is hardly affected by pressure. The 5$f$ orbital occupancy is well described by the intermediate coupling scheme.  
\end{abstract}

%% make title
%%%%%%%%%%%%%%%%%%%%%%%%%%%%%%%%%%%%%%%%%%%%%%%%%%%%%%%%%%%%%%%%%%%%%%%%%%

\maketitle

%% introduction
%%%%%%%%%%%%%%%%%%%%%%%%%%%%%%%%%%%%%%%%%%%%%%%%%%%%%%%%%%%%%%%%%%%%%%%%%%

%% actinides and its 5f states
\emph{Introduction.} Perhaps the actinides are the most fascinating, but the least understood elements in the Periodic Table. They manifest a plethora of interesting physical properties, such as intricate $P-T$ phase diagrams, low-symmetry crystal structures, multiple valence states, heavy-fermion features, and unconventional superconductivity, but only a few of them were studied by experiments or theoretical calculations~\cite{RevModPhys.81.235}. There is no doubt that the physical properties of the actinides are dominated by their electronic structures, specifically, status of their 5$f$ states. The 5$f$ states are usually correlated. There exists tricky interplay between Coulomb interaction, Hund's exchange, and spin-orbit coupling. The 5$f$ states are also Janus-faced, i.e, they can exhibit either localized or itinerant characters depending on their surroundings~\cite{PhysRevB.11.2740}. The complex nature of the 5$f$ states gives rise to extremely complicated electronic structures and unprecedentedly exotic physics.

%% Pu vs. the late actinides
Over the last decades, considerable attentions have been given to the light and middle actinides. Of particular interest is plutonium (Pu), which locates at the nexus of an unusual $\sim$ 40\% volume change that occurs in the actinides~\cite{albers:2001}. Clearly, the 5$f$ states of Pu, which go from being delocalized to localized, should be responsible for the dramatic volume change and the other anomalous properties~\cite{Joyce2006920,PhysRevLett.51.2418}. There have been extensive investigations concerning its magnetism~\cite{PhysRevB.72.054416,PhysRevLett.101.126403,Janoscheke:2015}, electronic structures~\cite{shim:2007,zhu:2013,savrasov:2001,PhysRevLett.101.056403,PhysRevB.75.235107,PhysRevB.82.085117}, lattice dynamics~\cite{dai:2003,PhysRevLett.92.146401,wong:2003}, phase transitions and phase stability~\cite{PhysRevLett.96.206402,PhysRevLett.92.185702,PhysRevB.81.224110,PhysRevX.5.011008}. On the contrary, the late actinides, such as americium (Am), curium (Cm), berkelium (Bk), and californium (Cf), have received much less attentions~\cite{PhysRevB.63.214101,PhysRevLett.96.036404,PhysRevLett.94.097002,Heathman110,PhysRevLett.41.42,PhysRevLett.44.1230,PhysRevLett.85.2961}. The reasons are twofold. On one hand, these materials are toxic and radioactive, which makes handling difficult and expensive. On the other hand, it is generally believed that their 5$f$ states are highly localized at ambient pressure, and scarcely contribute to the chemical bonding~\cite{PhysRevLett.85.2961,PhysRevLett.44.1230,PhysRevLett.41.42}. However, beyond the 5$f$ localized-itinerant transition, we argue that these elements could present many intriguing physical properties under pressure. A typical example is Cf, whose cubic phase could provide an ideal test-bed for investigating the complex behaviors of 5$f$ states. 

%% Introduction to Cf
To our knowledge, Cf is the heaviest actinide on which lattice structure studies can be performed at present~\cite{RevModPhys.81.235}. Recently, S.~Heathman \emph{et al.} studied the crystal structure of Cf up to 100~GPa by using both X-ray diffraction and theoretical calculations~\cite{PhysRevB.87.214111}. They observed four different crystallographic phases. At ambient condition, Cf presents a mixture of Cf-I (double hexagonal-close-packed) and Cf-II (face-centered-cubic) phases. Then the mixture gradually converts to the Cf-II phase under pressure. This transformation is completed at about 14~GPa. Upon additional pressure, the Cf-II phase converts to another mixture of Cf-III (face-centered orthorhombic) and Cf-IV (base-centered orthorhombic) phases, which emerges at about 35~GPa and is retained at least up to 100~GPa. Note that the volume collapse during the II-III transition is about 4.8\%, and that during the II-IV transition is about 15\%, which are attributed to the delocalization of $5f$ states again. Cf is also the only actinide to exhibit more than one valence (viz., divalent, intermediate valence, and trivalent states) at near to ambient conditions~\cite{RevModPhys.81.235}. Previously theoretical calculations about its $5f-6d$ promotion energy suggested that Cf falls into the boundary region between divalent and trivalent metallic bonding~\cite{JOHANSSON1978467,PhysRevB.19.6615}. Recent experiments confirmed that its divalent state is metastable, which will transform into the intermediate valence state, and then the trivalent state under moderate pressure~\cite{PhysRevB.87.214111}. 

%% Our approach and main results
In the present work, we employed a state-of-the-art first-principles many-body approach, namely a combination of the density functional theory and the single-site dynamical mean-field theory (dubbed as DFT + DMFT)~\cite{RevModPhys.68.13,RevModPhys.78.865}, to study the electronic structure of cubic Cf-II phase under pressure~\cite{calc_detail}. This approach has been widely used to study the electronic structures and related physical properties for Pu, Am, Cm, and many other actinide-based materials~\cite{dai:2003,shim:2007,zhu:2013,savrasov:2001,PhysRevLett.101.056403,PhysRevB.75.235107,PhysRevB.82.085117,PhysRevLett.101.126403,Janoscheke:2015}. Our results suggest that external pressure can tune the 5$f$ states in a subtle manner. Several fascinating effects are identified in cubic phase Cf, such as the orbital-selective 5$f$ localized-itinerant transition, divalent-trivalent transition, restricted valence state fluctuation, and collapse of local magnetic moment.

%% Results
%%%%%%%%%%%%%%%%%%%%%%%%%%%%%%%%%%%%%%%%%%%%%%%%%%%%%%%%%%%%%%%%%%%%%%%%%%

\begin{figure*}[ht]
\centering
\includegraphics[width=\textwidth]{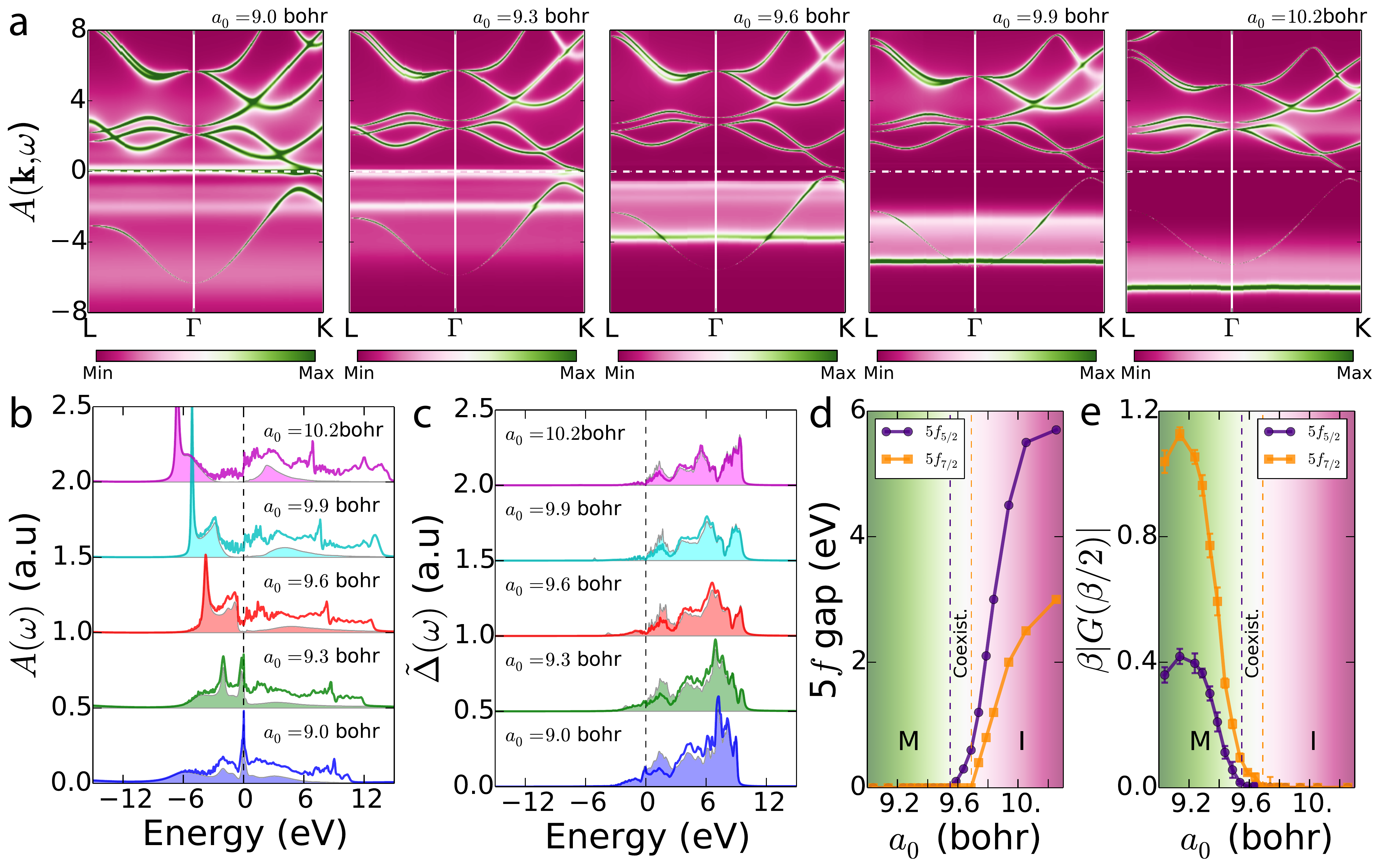}
\caption{(Color online). Pressure-dependent electronic structures of cubic phase Cf. (a) Momentum-resolved spectral functions $A(\mathbf{k},\omega)$. The horizontal lines denote the Fermi level. (b) Total density of states $A(\omega)$ (thick solid lines) and 5$f$ partial density of states $A_{5f}(\omega)$ (color-filled regions). (c) Imaginary parts of hybridization functions $\tilde{\Delta}(\omega) = -\text{Im} \Delta(\omega) / \pi$. The $5f_{5/2}$ and $5f_{7/2}$ components are represented by thick solid lines and color-filled regions, respectively. (d) The 5$f$ band gap as a function of lattice constants. (e) $\beta |G(\tau=\beta/2)|$ as a function of lattice constants, where $G(\tau)$ is the imaginary-time Green's function for 5$f$ orbitals and $\beta = 1/T$ is the inverse temperature. In panels (d) and (e), the letters ``M'' and ``I'' mean metallic and insulating characters, respectively. 
\label{fig:akw}}
\end{figure*}

\begin{figure*}[ht]
\centering
\includegraphics[width=\textwidth]{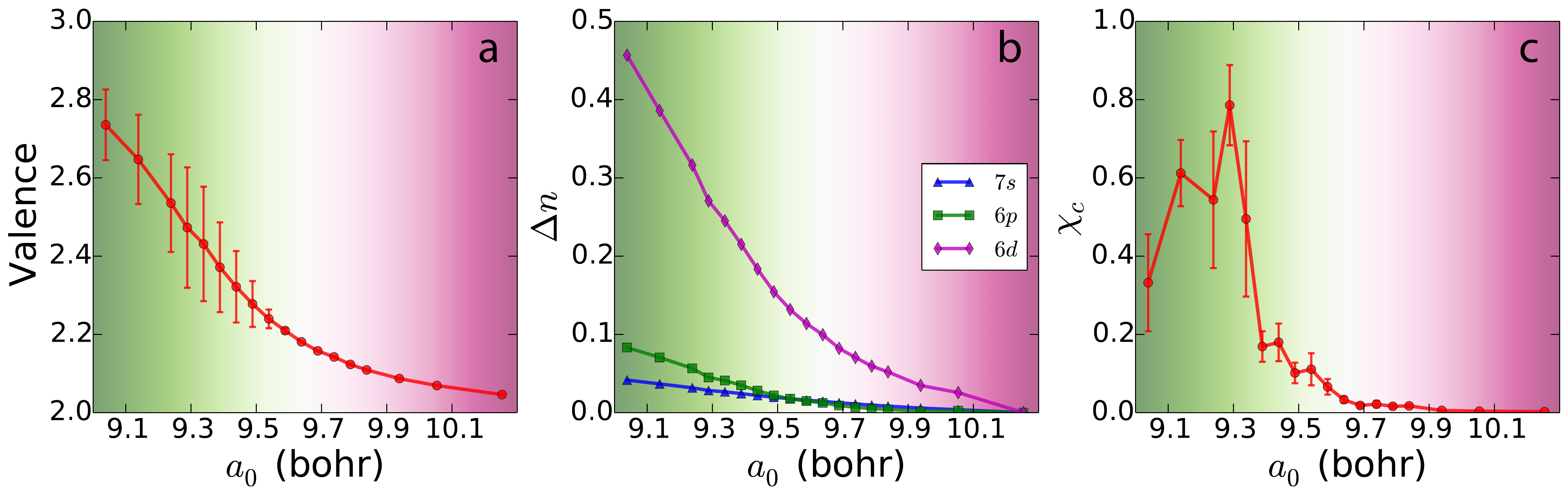}
\caption{(Color online). Valence state transition in cubic phase Cf upon volume compression. (a) Valence. The data are evaluated via the formula: valence = $12 - n_{5f}$, where $n_{5f}$ is the 5$f$ orbital occupancy. (b) Number of electrons those are promoted from $5f$ to $spd$ orbitals. (c) Charge fluctuation $\chi_c = \left(\langle n^2_{5f}\rangle - \langle n_{5f} \rangle^2 \right) / \beta$.  
\label{fig:val}}
\end{figure*}

\begin{figure*}[ht]
\centering
\includegraphics[width=\textwidth]{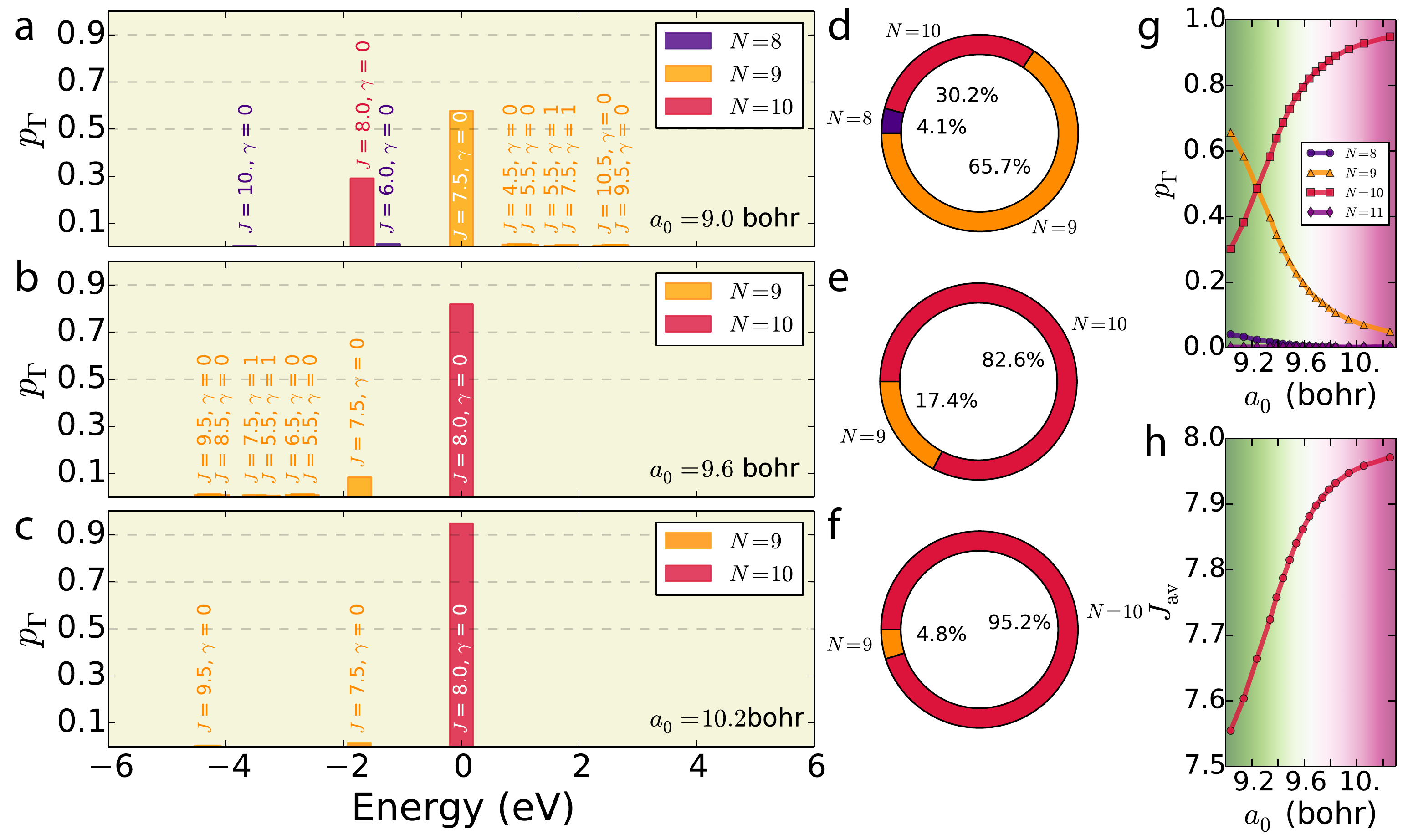}
\caption{(Color online). Valence state fluctuation in cubic phase Cf upon volume compression. (a)-(c) Probabilities of the 5$f$ atomic eigenstates (or equivalently valence state histograms) for various lattice volumes. (d)-(f) Distributions of 5$f$ atomic eigenstates with respect to $N$ for various lattice volumes. In these panels, only the contributions from the $N = 8$, $9$, and $10$ atomic eigenstates are shown. The contributions from the other atomic eigenstates are too trivial to be seen. (g) Distributions of 5$f$ atomic eigenstates with respect to volume compression. (h) Averaged total angular momentum $J_{\text{av}}$ as a function of lattice constants.
\label{fig:prob}}
\end{figure*}

\begin{figure}[ht]
\centering
\includegraphics[width=\columnwidth]{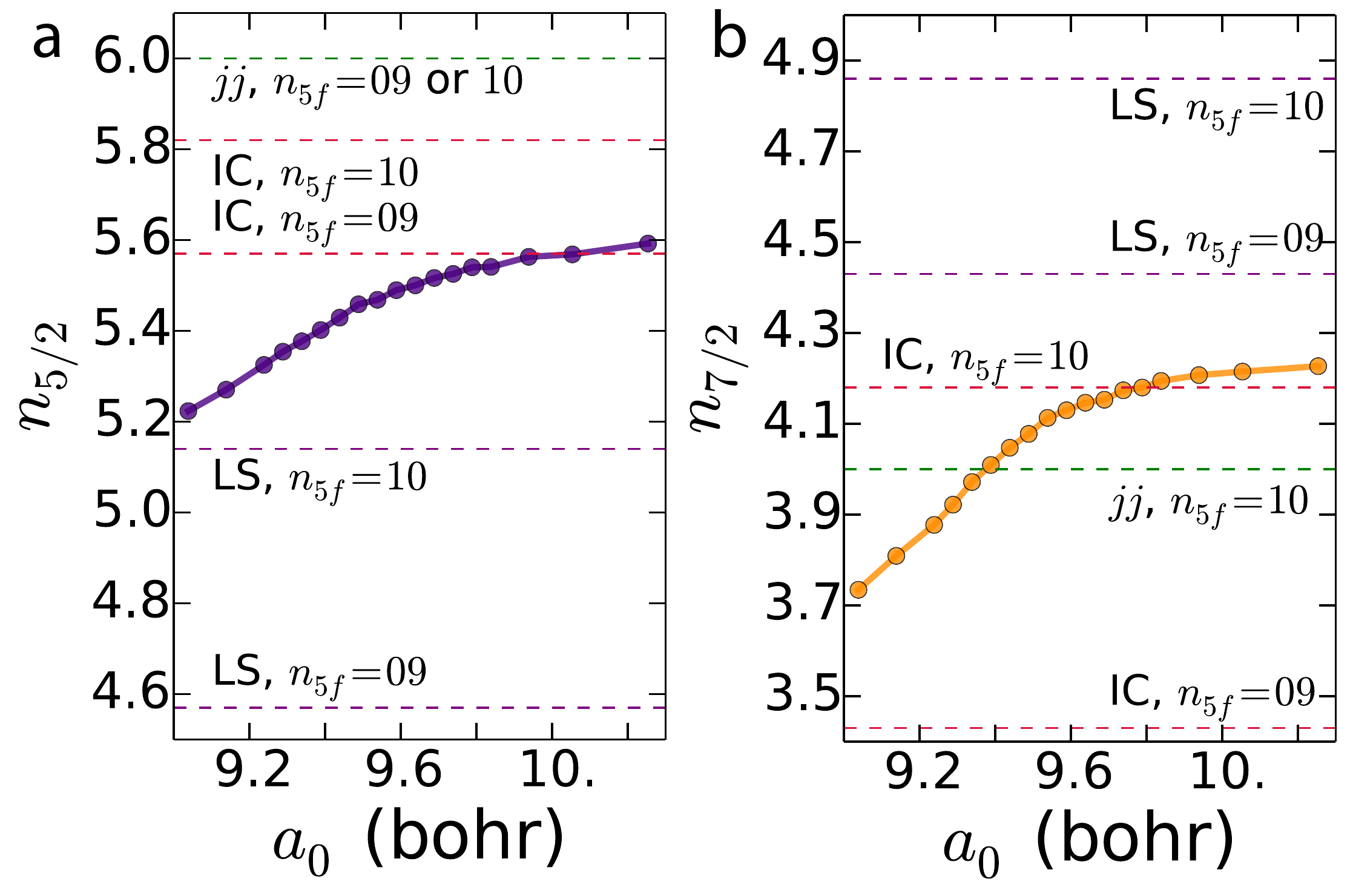}
\caption{(Color online). $5f$ orbital occupancy of cubic phase Cf upon volume compression. (a) $n_{5/2}$ case. (b) $n_{7/2}$ case. The horizontal dashed lines denote the theoretical values deduced by using various angular momentum coupling schemes (i.e, LS, IC, and $jj$)~\cite{RevModPhys.81.235}. As $n_{5f} = 09$ and the $jj$-coupling scheme is adopted, the theoretical value for $n_{7/2}$ is 3, which is not shown in panel (b).  
\label{fig:nimp}}
\end{figure}

\emph{Orbital-selective 5$f$ localized-itinerant transition.} Firstly, we calculate the spectral functions of cubic phase Cf with respect to various pressures (or lattice constants $a_0$). We concentrate on the momentum-resolved spectral functions $A(\mathbf{k},\omega)$, the total density of states $A(\omega)$, and the 5$f$ partial density of states $A_{5f}(\omega)$, which are depicted in Fig.~\ref{fig:akw}(a) and (b), respectively. The results endorse the scenario of pressure-driven electronic Lifshitz transition, which could be divided into three different stages. (i) $a_0 > 9.6$~bohr. There are stripe-like band structures in the momentum-resolved spectral functions, which are associated with the 5$f$ states. These bands are nearly flat and far away from the Fermi level, indicating the localized nature of the 5$f$ states. The 5$f$ partial density of states show obviously insulating behavior. When $a_0 = 10.2$~bohr, the 5$f$ band gap is estimated to be 3.0~eV. (ii) $a_0 \sim 9.6$~bohr. The occupied 5$f$ bands are shifted toward the Fermi level. In consequence, the 5$f$ band gap is greatly reduced. (iii) $a_0 < 9.6$~bohr. Strong quasi-particle peak emerges at the Fermi level, which is attributed to the itinerant 5$f$ states. The 5$f$ band gap is completely closed. We thus speculate that there is a small-to-large Fermi surface transition, accompanied by change of the Fermi surface topology~\cite{shim:1615,PhysRevLett.108.016402}. 

Then we focus on the hybridization functions $\Delta(\omega)$, which are generally used to measure the hybridization strength between the correlated $5f$ and non-correlated $spd$ bands. Fig.~\ref{fig:akw}(c) depicts the imaginary parts of hybridization functions. The $5f-spd$ hybridization mostly takes place at the unoccupied states. However, when $a_0 < 9.6$~bohr, there is sizeable spectral weight transferring from unoccupied to occupied states. All these features suggest that under pressure, the 5$f$ states in Cf should undergo a typical localized-itinerant transition. According to the experimental $P-V$ curve~\cite{PhysRevB.87.214111}, the transition will happen around 10~GPa. Note that similar transitions have been already observed at the high-pressure phases of Am and Cm~\cite{Heathman110,PhysRevLett.85.2961,PhysRevLett.41.42,PhysRevLett.44.1230}. 

Let's make a further analysis about this localized-itinerant transition. Due to the spin-orbit coupling, the 5$f$ orbitals can be split into the $5f_{5/2}$ and $5f_{7/2}$ sub-bands. The most astonished thing is that the localized-itinerant transitions don't occur simultaneously for the $5f_{5/2}$ and $5f_{7/2}$ states. In Fig.~\ref{fig:akw}(d), the 5$f$ band gap as a function of lattice constants is plotted. When $a_0 < 9.6$~bohr (or $a_0 > 9.7$~bohr), both the $5f_{5/2}$ and $5f_{7/2}$ states are metallic (or insulating). However, at the intermediate regime (i.e, $9.6$~bohr $< a_0< $ $9.7$~bohr), the $5f_{7/2}$ states become metallic while the $5f_{5/2}$ states still retain insulating. In order to confirm the coexistent zone, we further calculate $\beta |G(\tau=\beta/2)|$, which is in proportion to $A(\omega = 0)$ when $\beta \to \infty$~\cite{EPL.84.37009}. The calculated results are quite similar, as is seen in Fig.~\ref{fig:akw}(e). Therefore, we believe that this localized-itinerant transition is orbital-selective, which is an analogy to the orbital-selective Mott insulator-metal transition for multi-orbital correlated electron systems~\cite{PhysRevLett.92.216402,PhysRevB.85.245110}.

\emph{Valence state transition and charge redistribution.} As mentioned before, a valence state transition (divalent-trivalent) would take place upon volume compression~\cite{PhysRevB.87.214111,JOHANSSON1978467,PhysRevB.19.6615}. Previous studies suggested that the low $5f-6d$ promotion energy will facilitate this transition, but the underlying mechanism remains unclear so far. In the present work, we investigate the pressure-driven charge fluctuation, and find some useful clues. Firstly, even if the structural transition is ignored and only the Cf-II phase is considered, we can reproduce the valence state transition [see Fig.~\ref{fig:val}(a)]. Second, the 5$f$ electrons are mainly promoted to the $6d$ orbitals, but the electron transfer from $5f$ to $6p$ (or 7s) orbitals can not be neglected as well [see Fig.~\ref{fig:val}(b)]. Third, the charge fluctuation $\chi_c$ reaches its local maxima at $a_0 \sim 9.3$ bohr [see Fig.~\ref{fig:val}(c)]. Growing $5f-spd$ hybridization and the crossover of $5f^{10}$ and $5f^{9}$ levels may explain this peak. Since $\chi_c$ is proportional to the system's compressibility~\cite{RevModPhys.68.13}, it is reasonable to predict that this maxima leads to softening of the entire electronic liquid, which will manifest in the $P-V$ curve of Cf, in analogy to the fluctuating valence metal Yb~\cite{PhysRevLett.102.246401}. 

\emph{Valence state transition and atomic eigenvalues fluctuation.} The valence state histogram $p_{\Gamma}$ represents the probability to find a valence electron in a given atomic eigenstate $|\psi_{\Gamma}\rangle$, which is in general labelled by assorted good quantum numbers, such as $N$ (occupancy), $J$ (total angular momentum), and $\gamma$ (it stands for the rest good quantum numbers). It is a versatile tool to study the electronic configurations of realistic materials~\cite{shim:2007}. In Fig.~\ref{fig:prob}(a)-(c), the valence state histograms for three typical cases ($a_0 = 9.0$, $9.6$, and $10.2$~bohr) are shown. The corresponding distributions of 5$f$ atomic eigenstates with respect to $N$ are illustrated in Fig.~\ref{fig:prob}(d)-(f). When $a_0 = 10.2$~bohr, the atomic eigenvalues fluctuation is very weak. The 5$f$ electrons are virtually locked into the $5f^{10}$ configuration, and the valence state histogram is peaked only at the ground state of the atom (i.e, $|N = 10, J = 8, \gamma = 0 \rangle$). Cf behaves like a divalent metal (valence $\sim 2.0$, $n_{5f} \sim 10.0$). When $a_0 = 9.6$~bohr, the atomic eigenvalues fluctuation gets to be significant. Though the $5f^{10}$ configuration remains predominant, the contributions from the $5f^{9}$ configuration are not trivial. As a result, Cf begins to exhibit mixed-valence properties (valence $\sim 2.2$, $n_{5f} \sim 9.8$). When $a_0 = 9.0$~bohr, the dominant atomic eigenstate is $|N = 9, J = 7.5, \gamma = 0\rangle$, which is also the ground state of the atom. The atomic eigenstate $|N = 10, J = 8, \gamma = 0 \rangle$ becomes less important. It appears that the 5$f$ electrons live a double life, spending nearly all the time in the two states. In other words, the atomic eigenstates fluctuation only involves two principal states. The other atomic eigenstates are practically excluded. We call this behavior as ``restricted atomic eigenstate fluctuation'', which distinguishes Cf from the other typical mixed-valence materials, such as Pu~\cite{shim:2007,Janoscheke:2015}. At this moment, Cf is nearly a trivalent metal (valence $\sim 2.7$, $n_{5f} \sim 9.3$). Fig.~\ref{fig:prob}(g) shows the probabilities for the $5f^{8}-5f^{11}$ configurations. The probabilities for the $5f^{10}$ configuration increase monotonously against the lattice constants, while the trends for the other configurations are opposite. Besides, we find that there is a cross for the $5f^{10}$ and $5f^{9}$ configurations at $a_0 \sim 9.3$~bohr, which is related to the local maxima spotted in the local charge fluctuation $\chi_c$. Accordingly, we believe that the charge redistribution and atomic eigenstates fluctuation are probably the driving forces for the divalent-trivalent transition. 

Apart from the valence and the 5$f$ occupancy, it is also helpful to study the evolution of the other physical quantities with respect to the lattice constants. We calculate the averaged total angular momentum $J_{\text{av}}$ via the following equation: $J_{\text{av}} = \sum_{\Gamma} p_{\Gamma}J_{\Gamma}$, where $J_{\Gamma}$ denotes the total angular momentum for the atomic eigenstate $|\psi_{\Gamma}\rangle$. The results are plotted at Fig.~\ref{fig:prob}(h). As is expected, we observe sizeable collapse for $J_{\text{av}}$ under pressure. Note that $J_{\text{av}}$ declines more quickly when $a_0 < 9.6$~bohr. Since $J_{\text{av}}$ is tightly connected with the local magnetic moment $\mu$. It is plausible to suspect that $\mu$ should exhibit the same tendency, which is similar to the high-spin to low-spin transition in multi-orbital Mott systems~\cite{PhysRevLett.99.126405}. 

\emph{5$f$ orbital occupancy and angular momentum coupling.} As is well known, there are two standard ways to couple the angular momenta of multi-electronic systems: Russell-Saunders (LS) and $jj$ coupling. Provided that the spin-orbit coupling is weak compared to the electrostatic interactions, the LS coupling is favourable, or else the $jj$ coupling wins~\cite{RevModPhys.81.235}. As for the actinides, however, both the spin-orbit coupling and the electrostatic interactions are important. In this regard, the intermediate coupling (IC) scheme which takes both interactions into account is more appropriate~\cite{shim_epl09}. One exception to this rule is uranium, which exhibits LS coupling. What's the angular momentum coupling scheme for Cf? Will it be changed under pressure? In order to answer these questions, we calculate its 5$f$ orbital occupancy (see Fig.~\ref{fig:nimp}). The calculated values obviously favor the IC scheme. For example, given that $n_{5f} \sim 10.0$ ($a_0 = 10.2$~bohr), the calculated $n_{7/2}$ is approximately 4.23~\cite{occupancy}. However, the theoretical occupation numbers for $5f_{7/2}$ states are 4.86, 4.0, and 4.18 by LS, $jj$, and IC coupling schemes, respectively. The corresponding errors are 14.9\%, -5.4\%, and -1.2\%, respectively. Hence, we infer that the IC scheme still holds for Cf at ambient condition. Furthermore, the angular momentum coupling scheme won't be changed over the entire pressure (or volume) range of interest. Because the 5$f$ orbital occupancy can be used to calculate the X-ray branching ratio, the electron energy-loss spectroscopy and X-ray absorption spectroscopy can be employed to validate our results.  

%% Summary
%%%%%%%%%%%%%%%%%%%%%%%%%%%%%%%%%%%%%%%%%%%%%%%%%%%%%%%%%%%%%%%%%%%%%%%%%%

\emph{Conclusions.} We have examined the pressure-driven 5$f$ localized-itinerant transition and valence fluctuation in cubic phase Cf by using the DFT + DMFT approach. We believe that the $5f$ localized-itinerant transition is orbital-selective. There exists a considerable volume range where the insulating $5f_{5/2}$ state and the metallic $5f_{7/2}$ state could coexist. We also interpret the divalent-trivalent transition as a result of charge redistribution and valence state fluctuation. The valence state fluctuation is greatly restricted. It mainly involved two atomic eigenstates, in contrast to common fluctuating valence materials. Perhaps most importantly we confirm that Cf still obeys the IC scheme, which is not affected by pressure. Our results reveal that the $5f$ electronic structures for the late actinides under pressure are very interesting. Further theoretical and experimental investigations are highly desired.  

\begin{acknowledgments}
This work was supported by the Natural Science Foundation of China (No.~11504340 and 11704347), the Foundation of President of China Academy of Engineering Physics (No.~YZ2015012), and the Science Challenge Project of China (No.~TZ2016004).
\end{acknowledgments}

%% reference
%%%%%%%%%%%%%%%%%%%%%%%%%%%%%%%%%%%%%%%%%%%%%%%%%%%%%%%%%%%%%%%%%%%%%%%%%%

\bibliography{cf}

%merlin.mbs apsrev4-1.bst 2010-07-25 4.21a (PWD, AO, DPC) hacked
%Control: key (0)
%Control: author (8) initials jnrlst
%Control: editor formatted (1) identically to author
%Control: production of article title (-1) disabled
%Control: page (0) single
%Control: year (1) truncated
%Control: production of eprint (0) enabled
\begin{thebibliography}{43}%
\makeatletter
\providecommand \@ifxundefined [1]{%
 \@ifx{#1\undefined}
}%
\providecommand \@ifnum [1]{%
 \ifnum #1\expandafter \@firstoftwo
 \else \expandafter \@secondoftwo
 \fi
}%
\providecommand \@ifx [1]{%
 \ifx #1\expandafter \@firstoftwo
 \else \expandafter \@secondoftwo
 \fi
}%
\providecommand \natexlab [1]{#1}%
\providecommand \enquote  [1]{``#1''}%
\providecommand \bibnamefont  [1]{#1}%
\providecommand \bibfnamefont [1]{#1}%
\providecommand \citenamefont [1]{#1}%
\providecommand \href@noop [0]{\@secondoftwo}%
\providecommand \href [0]{\begingroup \@sanitize@url \@href}%
\providecommand \@href[1]{\@@startlink{#1}\@@href}%
\providecommand \@@href[1]{\endgroup#1\@@endlink}%
\providecommand \@sanitize@url [0]{\catcode `\\12\catcode `\$12\catcode
  `\&12\catcode `\#12\catcode `\^12\catcode `\_12\catcode `\%12\relax}%
\providecommand \@@startlink[1]{}%
\providecommand \@@endlink[0]{}%
\providecommand \url  [0]{\begingroup\@sanitize@url \@url }%
\providecommand \@url [1]{\endgroup\@href {#1}{\urlprefix }}%
\providecommand \urlprefix  [0]{URL }%
\providecommand \Eprint [0]{\href }%
\providecommand \doibase [0]{http://dx.doi.org/}%
\providecommand \selectlanguage [0]{\@gobble}%
\providecommand \bibinfo  [0]{\@secondoftwo}%
\providecommand \bibfield  [0]{\@secondoftwo}%
\providecommand \translation [1]{[#1]}%
\providecommand \BibitemOpen [0]{}%
\providecommand \bibitemStop [0]{}%
\providecommand \bibitemNoStop [0]{.\EOS\space}%
\providecommand \EOS [0]{\spacefactor3000\relax}%
\providecommand \BibitemShut  [1]{\csname bibitem#1\endcsname}%
\let\auto@bib@innerbib\@empty
%</preamble>
\bibitem [{\citenamefont {Moore}\ and\ \citenamefont {van~der
  Laan}(2009)}]{RevModPhys.81.235}%
  \BibitemOpen
  \bibfield  {author} {\bibinfo {author} {\bibfnamefont {K.~T.}\ \bibnamefont
  {Moore}}\ and\ \bibinfo {author} {\bibfnamefont {G.}~\bibnamefont {van~der
  Laan}},\ }\href {\doibase 10.1103/RevModPhys.81.235} {\bibfield  {journal}
  {\bibinfo  {journal} {Rev. Mod. Phys.}\ }\textbf {\bibinfo {volume} {81}},\
  \bibinfo {pages} {235} (\bibinfo {year} {2009})}\BibitemShut {NoStop}%
\bibitem [{\citenamefont {Johansson}(1975)}]{PhysRevB.11.2740}%
  \BibitemOpen
  \bibfield  {author} {\bibinfo {author} {\bibfnamefont {B.}~\bibnamefont
  {Johansson}},\ }\href {\doibase 10.1103/PhysRevB.11.2740} {\bibfield
  {journal} {\bibinfo  {journal} {Phys. Rev. B}\ }\textbf {\bibinfo {volume}
  {11}},\ \bibinfo {pages} {2740} (\bibinfo {year} {1975})}\BibitemShut
  {NoStop}%
\bibitem [{\citenamefont {Albers}(2001)}]{albers:2001}%
  \BibitemOpen
  \bibfield  {author} {\bibinfo {author} {\bibfnamefont {R.~C.}\ \bibnamefont
  {Albers}},\ }\href {\doibase 10.1038/35071205} {\bibfield  {journal}
  {\bibinfo  {journal} {Nature}\ }\textbf {\bibinfo {volume} {410}},\ \bibinfo
  {pages} {759} (\bibinfo {year} {2001})}\BibitemShut {NoStop}%
\bibitem [{\citenamefont {Joyce}\ \emph {et~al.}(2006)\citenamefont {Joyce},
  \citenamefont {Wills}, \citenamefont {Durakiewicz}, \citenamefont
  {Butterfield}, \citenamefont {Guziewicz}, \citenamefont {Moore},
  \citenamefont {Sarrao}, \citenamefont {Morales}, \citenamefont {Arko},
  \citenamefont {Eriksson}, \citenamefont {Delin},\ and\ \citenamefont
  {Graham}}]{Joyce2006920}%
  \BibitemOpen
  \bibfield  {author} {\bibinfo {author} {\bibfnamefont {J.}~\bibnamefont
  {Joyce}}, \bibinfo {author} {\bibfnamefont {J.}~\bibnamefont {Wills}},
  \bibinfo {author} {\bibfnamefont {T.}~\bibnamefont {Durakiewicz}}, \bibinfo
  {author} {\bibfnamefont {M.}~\bibnamefont {Butterfield}}, \bibinfo {author}
  {\bibfnamefont {E.}~\bibnamefont {Guziewicz}}, \bibinfo {author}
  {\bibfnamefont {D.}~\bibnamefont {Moore}}, \bibinfo {author} {\bibfnamefont
  {J.}~\bibnamefont {Sarrao}}, \bibinfo {author} {\bibfnamefont
  {L.}~\bibnamefont {Morales}}, \bibinfo {author} {\bibfnamefont
  {A.}~\bibnamefont {Arko}}, \bibinfo {author} {\bibfnamefont {O.}~\bibnamefont
  {Eriksson}}, \bibinfo {author} {\bibfnamefont {A.}~\bibnamefont {Delin}}, \
  and\ \bibinfo {author} {\bibfnamefont {K.}~\bibnamefont {Graham}},\ }\href
  {\doibase 10.1016/j.physb.2006.01.493} {\bibfield  {journal} {\bibinfo
  {journal} {Physica B: Condensed Matter}\ }\textbf {\bibinfo {volume}
  {378–380}},\ \bibinfo {pages} {920 } (\bibinfo {year} {2006})}\BibitemShut
  {NoStop}%
\bibitem [{\citenamefont {Cooper}\ \emph {et~al.}(1983)\citenamefont {Cooper},
  \citenamefont {Thayamballi}, \citenamefont {Spirlet}, \citenamefont
  {M\"uller},\ and\ \citenamefont {Vogt}}]{PhysRevLett.51.2418}%
  \BibitemOpen
  \bibfield  {author} {\bibinfo {author} {\bibfnamefont {B.~R.}\ \bibnamefont
  {Cooper}}, \bibinfo {author} {\bibfnamefont {P.}~\bibnamefont {Thayamballi}},
  \bibinfo {author} {\bibfnamefont {J.~C.}\ \bibnamefont {Spirlet}}, \bibinfo
  {author} {\bibfnamefont {W.}~\bibnamefont {M\"uller}}, \ and\ \bibinfo
  {author} {\bibfnamefont {O.}~\bibnamefont {Vogt}},\ }\href {\doibase
  10.1103/PhysRevLett.51.2418} {\bibfield  {journal} {\bibinfo  {journal}
  {Phys. Rev. Lett.}\ }\textbf {\bibinfo {volume} {51}},\ \bibinfo {pages}
  {2418} (\bibinfo {year} {1983})}\BibitemShut {NoStop}%
\bibitem [{\citenamefont {Lashley}\ \emph {et~al.}(2005)\citenamefont
  {Lashley}, \citenamefont {Lawson}, \citenamefont {McQueeney},\ and\
  \citenamefont {Lander}}]{PhysRevB.72.054416}%
  \BibitemOpen
  \bibfield  {author} {\bibinfo {author} {\bibfnamefont {J.~C.}\ \bibnamefont
  {Lashley}}, \bibinfo {author} {\bibfnamefont {A.}~\bibnamefont {Lawson}},
  \bibinfo {author} {\bibfnamefont {R.~J.}\ \bibnamefont {McQueeney}}, \ and\
  \bibinfo {author} {\bibfnamefont {G.~H.}\ \bibnamefont {Lander}},\ }\href
  {\doibase 10.1103/PhysRevB.72.054416} {\bibfield  {journal} {\bibinfo
  {journal} {Phys. Rev. B}\ }\textbf {\bibinfo {volume} {72}},\ \bibinfo
  {pages} {054416} (\bibinfo {year} {2005})}\BibitemShut {NoStop}%
\bibitem [{\citenamefont {Shim}\ \emph {et~al.}(2008)\citenamefont {Shim},
  \citenamefont {Haule}, \citenamefont {Savrasov},\ and\ \citenamefont
  {Kotliar}}]{PhysRevLett.101.126403}%
  \BibitemOpen
  \bibfield  {author} {\bibinfo {author} {\bibfnamefont {J.~H.}\ \bibnamefont
  {Shim}}, \bibinfo {author} {\bibfnamefont {K.}~\bibnamefont {Haule}},
  \bibinfo {author} {\bibfnamefont {S.}~\bibnamefont {Savrasov}}, \ and\
  \bibinfo {author} {\bibfnamefont {G.}~\bibnamefont {Kotliar}},\ }\href
  {\doibase 10.1103/PhysRevLett.101.126403} {\bibfield  {journal} {\bibinfo
  {journal} {Phys. Rev. Lett.}\ }\textbf {\bibinfo {volume} {101}},\ \bibinfo
  {pages} {126403} (\bibinfo {year} {2008})}\BibitemShut {NoStop}%
\bibitem [{\citenamefont {Janoschek}\ \emph {et~al.}(2015)\citenamefont
  {Janoschek}, \citenamefont {Das}, \citenamefont {Chakrabarti}, \citenamefont
  {Abernathy}, \citenamefont {Lumsden}, \citenamefont {Lawrence}, \citenamefont
  {Thompson}, \citenamefont {Lander}, \citenamefont {Mitchell}, \citenamefont
  {Richmond}, \citenamefont {Ramos}, \citenamefont {Trouw}, \citenamefont
  {Zhu}, \citenamefont {Haule}, \citenamefont {Kotliar},\ and\ \citenamefont
  {Bauer}}]{Janoscheke:2015}%
  \BibitemOpen
  \bibfield  {author} {\bibinfo {author} {\bibfnamefont {M.}~\bibnamefont
  {Janoschek}}, \bibinfo {author} {\bibfnamefont {P.}~\bibnamefont {Das}},
  \bibinfo {author} {\bibfnamefont {B.}~\bibnamefont {Chakrabarti}}, \bibinfo
  {author} {\bibfnamefont {D.~L.}\ \bibnamefont {Abernathy}}, \bibinfo {author}
  {\bibfnamefont {M.~D.}\ \bibnamefont {Lumsden}}, \bibinfo {author}
  {\bibfnamefont {J.~M.}\ \bibnamefont {Lawrence}}, \bibinfo {author}
  {\bibfnamefont {J.~D.}\ \bibnamefont {Thompson}}, \bibinfo {author}
  {\bibfnamefont {G.~H.}\ \bibnamefont {Lander}}, \bibinfo {author}
  {\bibfnamefont {J.~N.}\ \bibnamefont {Mitchell}}, \bibinfo {author}
  {\bibfnamefont {S.}~\bibnamefont {Richmond}}, \bibinfo {author}
  {\bibfnamefont {M.}~\bibnamefont {Ramos}}, \bibinfo {author} {\bibfnamefont
  {F.}~\bibnamefont {Trouw}}, \bibinfo {author} {\bibfnamefont {J.-X.}\
  \bibnamefont {Zhu}}, \bibinfo {author} {\bibfnamefont {K.}~\bibnamefont
  {Haule}}, \bibinfo {author} {\bibfnamefont {G.}~\bibnamefont {Kotliar}}, \
  and\ \bibinfo {author} {\bibfnamefont {E.~D.}\ \bibnamefont {Bauer}},\ }\href
  {\doibase 10.1126/sciadv.1500188} {\bibfield  {journal} {\bibinfo  {journal}
  {Sci. Adv.}\ }\textbf {\bibinfo {volume} {1}},\ \bibinfo {pages} {1500188}
  (\bibinfo {year} {2015})}\BibitemShut {NoStop}%
\bibitem [{\citenamefont {Shim}\ \emph
  {et~al.}(2007{\natexlab{a}})\citenamefont {Shim}, \citenamefont {Haule},\
  and\ \citenamefont {Kotliar}}]{shim:2007}%
  \BibitemOpen
  \bibfield  {author} {\bibinfo {author} {\bibfnamefont {J.~H.}\ \bibnamefont
  {Shim}}, \bibinfo {author} {\bibfnamefont {K.}~\bibnamefont {Haule}}, \ and\
  \bibinfo {author} {\bibfnamefont {G.}~\bibnamefont {Kotliar}},\ }\href
  {\doibase 10.1038/nature05647} {\bibfield  {journal} {\bibinfo  {journal}
  {Nature}\ }\textbf {\bibinfo {volume} {446}},\ \bibinfo {pages} {513}
  (\bibinfo {year} {2007}{\natexlab{a}})}\BibitemShut {NoStop}%
\bibitem [{\citenamefont {Zhu}\ \emph {et~al.}(2013)\citenamefont {Zhu},
  \citenamefont {Albers}, \citenamefont {Haule}, \citenamefont {Kotliar},\ and\
  \citenamefont {Wills}}]{zhu:2013}%
  \BibitemOpen
  \bibfield  {author} {\bibinfo {author} {\bibfnamefont {J.-X.}\ \bibnamefont
  {Zhu}}, \bibinfo {author} {\bibfnamefont {R.~C.}\ \bibnamefont {Albers}},
  \bibinfo {author} {\bibfnamefont {K.}~\bibnamefont {Haule}}, \bibinfo
  {author} {\bibfnamefont {G.}~\bibnamefont {Kotliar}}, \ and\ \bibinfo
  {author} {\bibfnamefont {J.~M.}\ \bibnamefont {Wills}},\ }\href {\doibase
  10.1038/ncomms3644} {\bibfield  {journal} {\bibinfo  {journal} {Nat Commun}\
  }\textbf {\bibinfo {volume} {4}},\ \bibinfo {pages} {3644} (\bibinfo {year}
  {2013})}\BibitemShut {NoStop}%
\bibitem [{\citenamefont {Savrasov}\ \emph {et~al.}(2001)\citenamefont
  {Savrasov}, \citenamefont {Kotliar},\ and\ \citenamefont
  {Abrahams}}]{savrasov:2001}%
  \BibitemOpen
  \bibfield  {author} {\bibinfo {author} {\bibfnamefont {S.~Y.}\ \bibnamefont
  {Savrasov}}, \bibinfo {author} {\bibfnamefont {G.}~\bibnamefont {Kotliar}}, \
  and\ \bibinfo {author} {\bibfnamefont {E.}~\bibnamefont {Abrahams}},\ }\href
  {\doibase 10.1038/35071035} {\bibfield  {journal} {\bibinfo  {journal}
  {Nature}\ }\textbf {\bibinfo {volume} {410}},\ \bibinfo {pages} {793}
  (\bibinfo {year} {2001})}\BibitemShut {NoStop}%
\bibitem [{\citenamefont {Marianetti}\ \emph {et~al.}(2008)\citenamefont
  {Marianetti}, \citenamefont {Haule}, \citenamefont {Kotliar},\ and\
  \citenamefont {Fluss}}]{PhysRevLett.101.056403}%
  \BibitemOpen
  \bibfield  {author} {\bibinfo {author} {\bibfnamefont {C.~A.}\ \bibnamefont
  {Marianetti}}, \bibinfo {author} {\bibfnamefont {K.}~\bibnamefont {Haule}},
  \bibinfo {author} {\bibfnamefont {G.}~\bibnamefont {Kotliar}}, \ and\
  \bibinfo {author} {\bibfnamefont {M.~J.}\ \bibnamefont {Fluss}},\ }\href
  {\doibase 10.1103/PhysRevLett.101.056403} {\bibfield  {journal} {\bibinfo
  {journal} {Phys. Rev. Lett.}\ }\textbf {\bibinfo {volume} {101}},\ \bibinfo
  {pages} {056403} (\bibinfo {year} {2008})}\BibitemShut {NoStop}%
\bibitem [{\citenamefont {Pourovskii}\ \emph {et~al.}(2007)\citenamefont
  {Pourovskii}, \citenamefont {Kotliar}, \citenamefont {Katsnelson},\ and\
  \citenamefont {Lichtenstein}}]{PhysRevB.75.235107}%
  \BibitemOpen
  \bibfield  {author} {\bibinfo {author} {\bibfnamefont {L.~V.}\ \bibnamefont
  {Pourovskii}}, \bibinfo {author} {\bibfnamefont {G.}~\bibnamefont {Kotliar}},
  \bibinfo {author} {\bibfnamefont {M.~I.}\ \bibnamefont {Katsnelson}}, \ and\
  \bibinfo {author} {\bibfnamefont {A.~I.}\ \bibnamefont {Lichtenstein}},\
  }\href {\doibase 10.1103/PhysRevB.75.235107} {\bibfield  {journal} {\bibinfo
  {journal} {Phys. Rev. B}\ }\textbf {\bibinfo {volume} {75}},\ \bibinfo
  {pages} {235107} (\bibinfo {year} {2007})}\BibitemShut {NoStop}%
\bibitem [{\citenamefont {Gorelov}\ \emph {et~al.}(2010)\citenamefont
  {Gorelov}, \citenamefont {Koloren\ifmmode~\check{c}\else \v{c}\fi{}},
  \citenamefont {Wehling}, \citenamefont {Hafermann}, \citenamefont {Shick},
  \citenamefont {Rubtsov}, \citenamefont {Landa}, \citenamefont {McMahan},
  \citenamefont {Anisimov}, \citenamefont {Katsnelson},\ and\ \citenamefont
  {Lichtenstein}}]{PhysRevB.82.085117}%
  \BibitemOpen
  \bibfield  {author} {\bibinfo {author} {\bibfnamefont {E.}~\bibnamefont
  {Gorelov}}, \bibinfo {author} {\bibfnamefont {J.}~\bibnamefont
  {Koloren\ifmmode~\check{c}\else \v{c}\fi{}}}, \bibinfo {author}
  {\bibfnamefont {T.}~\bibnamefont {Wehling}}, \bibinfo {author} {\bibfnamefont
  {H.}~\bibnamefont {Hafermann}}, \bibinfo {author} {\bibfnamefont {A.~B.}\
  \bibnamefont {Shick}}, \bibinfo {author} {\bibfnamefont {A.~N.}\ \bibnamefont
  {Rubtsov}}, \bibinfo {author} {\bibfnamefont {A.}~\bibnamefont {Landa}},
  \bibinfo {author} {\bibfnamefont {A.~K.}\ \bibnamefont {McMahan}}, \bibinfo
  {author} {\bibfnamefont {V.~I.}\ \bibnamefont {Anisimov}}, \bibinfo {author}
  {\bibfnamefont {M.~I.}\ \bibnamefont {Katsnelson}}, \ and\ \bibinfo {author}
  {\bibfnamefont {A.~I.}\ \bibnamefont {Lichtenstein}},\ }\href {\doibase
  10.1103/PhysRevB.82.085117} {\bibfield  {journal} {\bibinfo  {journal} {Phys.
  Rev. B}\ }\textbf {\bibinfo {volume} {82}},\ \bibinfo {pages} {085117}
  (\bibinfo {year} {2010})}\BibitemShut {NoStop}%
\bibitem [{\citenamefont {Dai}\ \emph {et~al.}(2003)\citenamefont {Dai},
  \citenamefont {Savrasov}, \citenamefont {Kotliar}, \citenamefont {Migliori},
  \citenamefont {Ledbetter},\ and\ \citenamefont {Abrahams}}]{dai:2003}%
  \BibitemOpen
  \bibfield  {author} {\bibinfo {author} {\bibfnamefont {X.}~\bibnamefont
  {Dai}}, \bibinfo {author} {\bibfnamefont {S.~Y.}\ \bibnamefont {Savrasov}},
  \bibinfo {author} {\bibfnamefont {G.}~\bibnamefont {Kotliar}}, \bibinfo
  {author} {\bibfnamefont {A.}~\bibnamefont {Migliori}}, \bibinfo {author}
  {\bibfnamefont {H.}~\bibnamefont {Ledbetter}}, \ and\ \bibinfo {author}
  {\bibfnamefont {E.}~\bibnamefont {Abrahams}},\ }\href {\doibase
  10.1126/science.1083428} {\bibfield  {journal} {\bibinfo  {journal}
  {Science}\ }\textbf {\bibinfo {volume} {300}},\ \bibinfo {pages} {953}
  (\bibinfo {year} {2003})}\BibitemShut {NoStop}%
\bibitem [{\citenamefont {McQueeney}\ \emph {et~al.}(2004)\citenamefont
  {McQueeney}, \citenamefont {Lawson}, \citenamefont {Migliori}, \citenamefont
  {Kelley}, \citenamefont {Fultz}, \citenamefont {Ramos}, \citenamefont
  {Martinez}, \citenamefont {Lashley},\ and\ \citenamefont
  {Vogel}}]{PhysRevLett.92.146401}%
  \BibitemOpen
  \bibfield  {author} {\bibinfo {author} {\bibfnamefont {R.~J.}\ \bibnamefont
  {McQueeney}}, \bibinfo {author} {\bibfnamefont {A.~C.}\ \bibnamefont
  {Lawson}}, \bibinfo {author} {\bibfnamefont {A.}~\bibnamefont {Migliori}},
  \bibinfo {author} {\bibfnamefont {T.~M.}\ \bibnamefont {Kelley}}, \bibinfo
  {author} {\bibfnamefont {B.}~\bibnamefont {Fultz}}, \bibinfo {author}
  {\bibfnamefont {M.}~\bibnamefont {Ramos}}, \bibinfo {author} {\bibfnamefont
  {B.}~\bibnamefont {Martinez}}, \bibinfo {author} {\bibfnamefont {J.~C.}\
  \bibnamefont {Lashley}}, \ and\ \bibinfo {author} {\bibfnamefont {S.~C.}\
  \bibnamefont {Vogel}},\ }\href {\doibase 10.1103/PhysRevLett.92.146401}
  {\bibfield  {journal} {\bibinfo  {journal} {Phys. Rev. Lett.}\ }\textbf
  {\bibinfo {volume} {92}},\ \bibinfo {pages} {146401} (\bibinfo {year}
  {2004})}\BibitemShut {NoStop}%
\bibitem [{\citenamefont {Wong}\ \emph {et~al.}(2003)\citenamefont {Wong},
  \citenamefont {Krisch}, \citenamefont {Farber}, \citenamefont {Occelli},
  \citenamefont {Schwartz}, \citenamefont {Chiang}, \citenamefont {Wall},
  \citenamefont {Boro},\ and\ \citenamefont {Xu}}]{wong:2003}%
  \BibitemOpen
  \bibfield  {author} {\bibinfo {author} {\bibfnamefont {J.}~\bibnamefont
  {Wong}}, \bibinfo {author} {\bibfnamefont {M.}~\bibnamefont {Krisch}},
  \bibinfo {author} {\bibfnamefont {D.~L.}\ \bibnamefont {Farber}}, \bibinfo
  {author} {\bibfnamefont {F.}~\bibnamefont {Occelli}}, \bibinfo {author}
  {\bibfnamefont {A.~J.}\ \bibnamefont {Schwartz}}, \bibinfo {author}
  {\bibfnamefont {T.-C.}\ \bibnamefont {Chiang}}, \bibinfo {author}
  {\bibfnamefont {M.}~\bibnamefont {Wall}}, \bibinfo {author} {\bibfnamefont
  {C.}~\bibnamefont {Boro}}, \ and\ \bibinfo {author} {\bibfnamefont
  {R.}~\bibnamefont {Xu}},\ }\href {\doibase 10.1126/science.1087179}
  {\bibfield  {journal} {\bibinfo  {journal} {Science}\ }\textbf {\bibinfo
  {volume} {301}},\ \bibinfo {pages} {1078} (\bibinfo {year}
  {2003})}\BibitemShut {NoStop}%
\bibitem [{\citenamefont {Moore}\ \emph {et~al.}(2006)\citenamefont {Moore},
  \citenamefont {S\"oderlind}, \citenamefont {Schwartz},\ and\ \citenamefont
  {Laughlin}}]{PhysRevLett.96.206402}%
  \BibitemOpen
  \bibfield  {author} {\bibinfo {author} {\bibfnamefont {K.~T.}\ \bibnamefont
  {Moore}}, \bibinfo {author} {\bibfnamefont {P.}~\bibnamefont {S\"oderlind}},
  \bibinfo {author} {\bibfnamefont {A.~J.}\ \bibnamefont {Schwartz}}, \ and\
  \bibinfo {author} {\bibfnamefont {D.~E.}\ \bibnamefont {Laughlin}},\ }\href
  {\doibase 10.1103/PhysRevLett.96.206402} {\bibfield  {journal} {\bibinfo
  {journal} {Phys. Rev. Lett.}\ }\textbf {\bibinfo {volume} {96}},\ \bibinfo
  {pages} {206402} (\bibinfo {year} {2006})}\BibitemShut {NoStop}%
\bibitem [{\citenamefont {S\"oderlind}\ and\ \citenamefont
  {Sadigh}(2004)}]{PhysRevLett.92.185702}%
  \BibitemOpen
  \bibfield  {author} {\bibinfo {author} {\bibfnamefont {P.}~\bibnamefont
  {S\"oderlind}}\ and\ \bibinfo {author} {\bibfnamefont {B.}~\bibnamefont
  {Sadigh}},\ }\href {\doibase 10.1103/PhysRevLett.92.185702} {\bibfield
  {journal} {\bibinfo  {journal} {Phys. Rev. Lett.}\ }\textbf {\bibinfo
  {volume} {92}},\ \bibinfo {pages} {185702} (\bibinfo {year}
  {2004})}\BibitemShut {NoStop}%
\bibitem [{\citenamefont {S\"oderlind}\ \emph {et~al.}(2010)\citenamefont
  {S\"oderlind}, \citenamefont {Landa}, \citenamefont {Klepeis}, \citenamefont
  {Suzuki},\ and\ \citenamefont {Migliori}}]{PhysRevB.81.224110}%
  \BibitemOpen
  \bibfield  {author} {\bibinfo {author} {\bibfnamefont {P.}~\bibnamefont
  {S\"oderlind}}, \bibinfo {author} {\bibfnamefont {A.}~\bibnamefont {Landa}},
  \bibinfo {author} {\bibfnamefont {J.~E.}\ \bibnamefont {Klepeis}}, \bibinfo
  {author} {\bibfnamefont {Y.}~\bibnamefont {Suzuki}}, \ and\ \bibinfo {author}
  {\bibfnamefont {A.}~\bibnamefont {Migliori}},\ }\href {\doibase
  10.1103/PhysRevB.81.224110} {\bibfield  {journal} {\bibinfo  {journal} {Phys.
  Rev. B}\ }\textbf {\bibinfo {volume} {81}},\ \bibinfo {pages} {224110}
  (\bibinfo {year} {2010})}\BibitemShut {NoStop}%
\bibitem [{\citenamefont {Lanat\`a}\ \emph {et~al.}(2015)\citenamefont
  {Lanat\`a}, \citenamefont {Yao}, \citenamefont {Wang}, \citenamefont {Ho},\
  and\ \citenamefont {Kotliar}}]{PhysRevX.5.011008}%
  \BibitemOpen
  \bibfield  {author} {\bibinfo {author} {\bibfnamefont {N.}~\bibnamefont
  {Lanat\`a}}, \bibinfo {author} {\bibfnamefont {Y.}~\bibnamefont {Yao}},
  \bibinfo {author} {\bibfnamefont {C.-Z.}\ \bibnamefont {Wang}}, \bibinfo
  {author} {\bibfnamefont {K.-M.}\ \bibnamefont {Ho}}, \ and\ \bibinfo {author}
  {\bibfnamefont {G.}~\bibnamefont {Kotliar}},\ }\href {\doibase
  10.1103/PhysRevX.5.011008} {\bibfield  {journal} {\bibinfo  {journal} {Phys.
  Rev. X}\ }\textbf {\bibinfo {volume} {5}},\ \bibinfo {pages} {011008}
  (\bibinfo {year} {2015})}\BibitemShut {NoStop}%
\bibitem [{\citenamefont {Lindbaum}\ \emph {et~al.}(2001)\citenamefont
  {Lindbaum}, \citenamefont {Heathman}, \citenamefont {Litfin}, \citenamefont
  {M\'eresse}, \citenamefont {Haire}, \citenamefont {Le~Bihan},\ and\
  \citenamefont {Libotte}}]{PhysRevB.63.214101}%
  \BibitemOpen
  \bibfield  {author} {\bibinfo {author} {\bibfnamefont {A.}~\bibnamefont
  {Lindbaum}}, \bibinfo {author} {\bibfnamefont {S.}~\bibnamefont {Heathman}},
  \bibinfo {author} {\bibfnamefont {K.}~\bibnamefont {Litfin}}, \bibinfo
  {author} {\bibfnamefont {Y.}~\bibnamefont {M\'eresse}}, \bibinfo {author}
  {\bibfnamefont {R.~G.}\ \bibnamefont {Haire}}, \bibinfo {author}
  {\bibfnamefont {T.}~\bibnamefont {Le~Bihan}}, \ and\ \bibinfo {author}
  {\bibfnamefont {H.}~\bibnamefont {Libotte}},\ }\href {\doibase
  10.1103/PhysRevB.63.214101} {\bibfield  {journal} {\bibinfo  {journal} {Phys.
  Rev. B}\ }\textbf {\bibinfo {volume} {63}},\ \bibinfo {pages} {214101}
  (\bibinfo {year} {2001})}\BibitemShut {NoStop}%
\bibitem [{\citenamefont {Savrasov}\ \emph {et~al.}(2006)\citenamefont
  {Savrasov}, \citenamefont {Haule},\ and\ \citenamefont
  {Kotliar}}]{PhysRevLett.96.036404}%
  \BibitemOpen
  \bibfield  {author} {\bibinfo {author} {\bibfnamefont {S.~Y.}\ \bibnamefont
  {Savrasov}}, \bibinfo {author} {\bibfnamefont {K.}~\bibnamefont {Haule}}, \
  and\ \bibinfo {author} {\bibfnamefont {G.}~\bibnamefont {Kotliar}},\ }\href
  {\doibase 10.1103/PhysRevLett.96.036404} {\bibfield  {journal} {\bibinfo
  {journal} {Phys. Rev. Lett.}\ }\textbf {\bibinfo {volume} {96}},\ \bibinfo
  {pages} {036404} (\bibinfo {year} {2006})}\BibitemShut {NoStop}%
\bibitem [{\citenamefont {Griveau}\ \emph {et~al.}(2005)\citenamefont
  {Griveau}, \citenamefont {Rebizant}, \citenamefont {Lander},\ and\
  \citenamefont {Kotliar}}]{PhysRevLett.94.097002}%
  \BibitemOpen
  \bibfield  {author} {\bibinfo {author} {\bibfnamefont {J.-C.}\ \bibnamefont
  {Griveau}}, \bibinfo {author} {\bibfnamefont {J.}~\bibnamefont {Rebizant}},
  \bibinfo {author} {\bibfnamefont {G.~H.}\ \bibnamefont {Lander}}, \ and\
  \bibinfo {author} {\bibfnamefont {G.}~\bibnamefont {Kotliar}},\ }\href
  {\doibase 10.1103/PhysRevLett.94.097002} {\bibfield  {journal} {\bibinfo
  {journal} {Phys. Rev. Lett.}\ }\textbf {\bibinfo {volume} {94}},\ \bibinfo
  {pages} {097002} (\bibinfo {year} {2005})}\BibitemShut {NoStop}%
\bibitem [{\citenamefont {Heathman}\ \emph {et~al.}(2005)\citenamefont
  {Heathman}, \citenamefont {Haire}, \citenamefont {Le~Bihan}, \citenamefont
  {Lindbaum}, \citenamefont {Idiri}, \citenamefont {Normile}, \citenamefont
  {Li}, \citenamefont {Ahuja}, \citenamefont {Johansson},\ and\ \citenamefont
  {Lander}}]{Heathman110}%
  \BibitemOpen
  \bibfield  {author} {\bibinfo {author} {\bibfnamefont {S.}~\bibnamefont
  {Heathman}}, \bibinfo {author} {\bibfnamefont {R.~G.}\ \bibnamefont {Haire}},
  \bibinfo {author} {\bibfnamefont {T.}~\bibnamefont {Le~Bihan}}, \bibinfo
  {author} {\bibfnamefont {A.}~\bibnamefont {Lindbaum}}, \bibinfo {author}
  {\bibfnamefont {M.}~\bibnamefont {Idiri}}, \bibinfo {author} {\bibfnamefont
  {P.}~\bibnamefont {Normile}}, \bibinfo {author} {\bibfnamefont
  {S.}~\bibnamefont {Li}}, \bibinfo {author} {\bibfnamefont {R.}~\bibnamefont
  {Ahuja}}, \bibinfo {author} {\bibfnamefont {B.}~\bibnamefont {Johansson}}, \
  and\ \bibinfo {author} {\bibfnamefont {G.~H.}\ \bibnamefont {Lander}},\
  }\href {\doibase 10.1126/science.1112453} {\bibfield  {journal} {\bibinfo
  {journal} {Science}\ }\textbf {\bibinfo {volume} {309}},\ \bibinfo {pages}
  {110} (\bibinfo {year} {2005})}\BibitemShut {NoStop}%
\bibitem [{\citenamefont {Skriver}\ \emph {et~al.}(1978)\citenamefont
  {Skriver}, \citenamefont {Andersen},\ and\ \citenamefont
  {Johansson}}]{PhysRevLett.41.42}%
  \BibitemOpen
  \bibfield  {author} {\bibinfo {author} {\bibfnamefont {H.~L.}\ \bibnamefont
  {Skriver}}, \bibinfo {author} {\bibfnamefont {O.~K.}\ \bibnamefont
  {Andersen}}, \ and\ \bibinfo {author} {\bibfnamefont {B.}~\bibnamefont
  {Johansson}},\ }\href {\doibase 10.1103/PhysRevLett.41.42} {\bibfield
  {journal} {\bibinfo  {journal} {Phys. Rev. Lett.}\ }\textbf {\bibinfo
  {volume} {41}},\ \bibinfo {pages} {42} (\bibinfo {year} {1978})}\BibitemShut
  {NoStop}%
\bibitem [{\citenamefont {Skriver}\ \emph {et~al.}(1980)\citenamefont
  {Skriver}, \citenamefont {Andersen},\ and\ \citenamefont
  {Johansson}}]{PhysRevLett.44.1230}%
  \BibitemOpen
  \bibfield  {author} {\bibinfo {author} {\bibfnamefont {H.~L.}\ \bibnamefont
  {Skriver}}, \bibinfo {author} {\bibfnamefont {O.~K.}\ \bibnamefont
  {Andersen}}, \ and\ \bibinfo {author} {\bibfnamefont {B.}~\bibnamefont
  {Johansson}},\ }\href {\doibase 10.1103/PhysRevLett.44.1230} {\bibfield
  {journal} {\bibinfo  {journal} {Phys. Rev. Lett.}\ }\textbf {\bibinfo
  {volume} {44}},\ \bibinfo {pages} {1230} (\bibinfo {year}
  {1980})}\BibitemShut {NoStop}%
\bibitem [{\citenamefont {Heathman}\ \emph {et~al.}(2000)\citenamefont
  {Heathman}, \citenamefont {Haire}, \citenamefont {Le~Bihan}, \citenamefont
  {Lindbaum}, \citenamefont {Litfin}, \citenamefont {M\'eresse},\ and\
  \citenamefont {Libotte}}]{PhysRevLett.85.2961}%
  \BibitemOpen
  \bibfield  {author} {\bibinfo {author} {\bibfnamefont {S.}~\bibnamefont
  {Heathman}}, \bibinfo {author} {\bibfnamefont {R.~G.}\ \bibnamefont {Haire}},
  \bibinfo {author} {\bibfnamefont {T.}~\bibnamefont {Le~Bihan}}, \bibinfo
  {author} {\bibfnamefont {A.}~\bibnamefont {Lindbaum}}, \bibinfo {author}
  {\bibfnamefont {K.}~\bibnamefont {Litfin}}, \bibinfo {author} {\bibfnamefont
  {Y.}~\bibnamefont {M\'eresse}}, \ and\ \bibinfo {author} {\bibfnamefont
  {H.}~\bibnamefont {Libotte}},\ }\href {\doibase 10.1103/PhysRevLett.85.2961}
  {\bibfield  {journal} {\bibinfo  {journal} {Phys. Rev. Lett.}\ }\textbf
  {\bibinfo {volume} {85}},\ \bibinfo {pages} {2961} (\bibinfo {year}
  {2000})}\BibitemShut {NoStop}%
\bibitem [{\citenamefont {Heathman}\ \emph {et~al.}(2013)\citenamefont
  {Heathman}, \citenamefont {Le~Bihan}, \citenamefont {Yagoubi}, \citenamefont
  {Johansson},\ and\ \citenamefont {Ahuja}}]{PhysRevB.87.214111}%
  \BibitemOpen
  \bibfield  {author} {\bibinfo {author} {\bibfnamefont {S.}~\bibnamefont
  {Heathman}}, \bibinfo {author} {\bibfnamefont {T.}~\bibnamefont {Le~Bihan}},
  \bibinfo {author} {\bibfnamefont {S.}~\bibnamefont {Yagoubi}}, \bibinfo
  {author} {\bibfnamefont {B.}~\bibnamefont {Johansson}}, \ and\ \bibinfo
  {author} {\bibfnamefont {R.}~\bibnamefont {Ahuja}},\ }\href {\doibase
  10.1103/PhysRevB.87.214111} {\bibfield  {journal} {\bibinfo  {journal} {Phys.
  Rev. B}\ }\textbf {\bibinfo {volume} {87}},\ \bibinfo {pages} {214111}
  (\bibinfo {year} {2013})}\BibitemShut {NoStop}%
\bibitem [{\citenamefont {Johansson}(1978)}]{JOHANSSON1978467}%
  \BibitemOpen
  \bibfield  {author} {\bibinfo {author} {\bibfnamefont {B.}~\bibnamefont
  {Johansson}},\ }\href {\doibase https://doi.org/10.1016/0022-3697(78)90023-9}
  {\bibfield  {journal} {\bibinfo  {journal} {J. Phys. Chem. Solids}\ }\textbf
  {\bibinfo {volume} {39}},\ \bibinfo {pages} {467 } (\bibinfo {year}
  {1978})}\BibitemShut {NoStop}%
\bibitem [{\citenamefont {Johansson}(1979)}]{PhysRevB.19.6615}%
  \BibitemOpen
  \bibfield  {author} {\bibinfo {author} {\bibfnamefont {B.}~\bibnamefont
  {Johansson}},\ }\href {\doibase 10.1103/PhysRevB.19.6615} {\bibfield
  {journal} {\bibinfo  {journal} {Phys. Rev. B}\ }\textbf {\bibinfo {volume}
  {19}},\ \bibinfo {pages} {6615} (\bibinfo {year} {1979})}\BibitemShut
  {NoStop}%
\bibitem [{\citenamefont {Georges}\ \emph {et~al.}(1996)\citenamefont
  {Georges}, \citenamefont {Kotliar}, \citenamefont {Krauth},\ and\
  \citenamefont {Rozenberg}}]{RevModPhys.68.13}%
  \BibitemOpen
  \bibfield  {author} {\bibinfo {author} {\bibfnamefont {A.}~\bibnamefont
  {Georges}}, \bibinfo {author} {\bibfnamefont {G.}~\bibnamefont {Kotliar}},
  \bibinfo {author} {\bibfnamefont {W.}~\bibnamefont {Krauth}}, \ and\ \bibinfo
  {author} {\bibfnamefont {M.~J.}\ \bibnamefont {Rozenberg}},\ }\href {\doibase
  10.1103/RevModPhys.68.13} {\bibfield  {journal} {\bibinfo  {journal} {Rev.
  Mod. Phys.}\ }\textbf {\bibinfo {volume} {68}},\ \bibinfo {pages} {13}
  (\bibinfo {year} {1996})}\BibitemShut {NoStop}%
\bibitem [{\citenamefont {Kotliar}\ \emph {et~al.}(2006)\citenamefont
  {Kotliar}, \citenamefont {Savrasov}, \citenamefont {Haule}, \citenamefont
  {Oudovenko}, \citenamefont {Parcollet},\ and\ \citenamefont
  {Marianetti}}]{RevModPhys.78.865}%
  \BibitemOpen
  \bibfield  {author} {\bibinfo {author} {\bibfnamefont {G.}~\bibnamefont
  {Kotliar}}, \bibinfo {author} {\bibfnamefont {S.~Y.}\ \bibnamefont
  {Savrasov}}, \bibinfo {author} {\bibfnamefont {K.}~\bibnamefont {Haule}},
  \bibinfo {author} {\bibfnamefont {V.~S.}\ \bibnamefont {Oudovenko}}, \bibinfo
  {author} {\bibfnamefont {O.}~\bibnamefont {Parcollet}}, \ and\ \bibinfo
  {author} {\bibfnamefont {C.~A.}\ \bibnamefont {Marianetti}},\ }\href
  {\doibase 10.1103/RevModPhys.78.865} {\bibfield  {journal} {\bibinfo
  {journal} {Rev. Mod. Phys.}\ }\textbf {\bibinfo {volume} {78}},\ \bibinfo
  {pages} {865} (\bibinfo {year} {2006})}\BibitemShut {NoStop}%
\bibitem [{cal()}]{calc_detail}%
  \BibitemOpen
  \href@noop {} {}\bibinfo {note} {See Supplementary Materials for
  computational details.}\BibitemShut {Stop}%
\bibitem [{\citenamefont {Shim}\ \emph
  {et~al.}(2007{\natexlab{b}})\citenamefont {Shim}, \citenamefont {Haule},\
  and\ \citenamefont {Kotliar}}]{shim:1615}%
  \BibitemOpen
  \bibfield  {author} {\bibinfo {author} {\bibfnamefont {J.~H.}\ \bibnamefont
  {Shim}}, \bibinfo {author} {\bibfnamefont {K.}~\bibnamefont {Haule}}, \ and\
  \bibinfo {author} {\bibfnamefont {G.}~\bibnamefont {Kotliar}},\ }\href
  {\doibase 10.1126/science.1149064} {\bibfield  {journal} {\bibinfo  {journal}
  {Science}\ }\textbf {\bibinfo {volume} {318}},\ \bibinfo {pages} {1615}
  (\bibinfo {year} {2007}{\natexlab{b}})}\BibitemShut {NoStop}%
\bibitem [{\citenamefont {Choi}\ \emph {et~al.}(2012)\citenamefont {Choi},
  \citenamefont {Min}, \citenamefont {Shim}, \citenamefont {Haule},\ and\
  \citenamefont {Kotliar}}]{PhysRevLett.108.016402}%
  \BibitemOpen
  \bibfield  {author} {\bibinfo {author} {\bibfnamefont {H.~C.}\ \bibnamefont
  {Choi}}, \bibinfo {author} {\bibfnamefont {B.~I.}\ \bibnamefont {Min}},
  \bibinfo {author} {\bibfnamefont {J.~H.}\ \bibnamefont {Shim}}, \bibinfo
  {author} {\bibfnamefont {K.}~\bibnamefont {Haule}}, \ and\ \bibinfo {author}
  {\bibfnamefont {G.}~\bibnamefont {Kotliar}},\ }\href {\doibase
  10.1103/PhysRevLett.108.016402} {\bibfield  {journal} {\bibinfo  {journal}
  {Phys. Rev. Lett.}\ }\textbf {\bibinfo {volume} {108}},\ \bibinfo {pages}
  {016402} (\bibinfo {year} {2012})}\BibitemShut {NoStop}%
\bibitem [{\citenamefont {Gull}\ \emph {et~al.}(2008)\citenamefont {Gull},
  \citenamefont {Werner}, \citenamefont {Wang}, \citenamefont {Troyer},\ and\
  \citenamefont {Millis}}]{EPL.84.37009}%
  \BibitemOpen
  \bibfield  {author} {\bibinfo {author} {\bibfnamefont {E.}~\bibnamefont
  {Gull}}, \bibinfo {author} {\bibfnamefont {P.}~\bibnamefont {Werner}},
  \bibinfo {author} {\bibfnamefont {X.}~\bibnamefont {Wang}}, \bibinfo {author}
  {\bibfnamefont {M.}~\bibnamefont {Troyer}}, \ and\ \bibinfo {author}
  {\bibfnamefont {A.~J.}\ \bibnamefont {Millis}},\ }\href
  {http://stacks.iop.org/0295-5075/84/i=3/a=37009} {\bibfield  {journal}
  {\bibinfo  {journal} {Europhys. Lett.}\ }\textbf {\bibinfo {volume} {84}},\
  \bibinfo {pages} {37009} (\bibinfo {year} {2008})}\BibitemShut {NoStop}%
\bibitem [{\citenamefont {Koga}\ \emph {et~al.}(2004)\citenamefont {Koga},
  \citenamefont {Kawakami}, \citenamefont {Rice},\ and\ \citenamefont
  {Sigrist}}]{PhysRevLett.92.216402}%
  \BibitemOpen
  \bibfield  {author} {\bibinfo {author} {\bibfnamefont {A.}~\bibnamefont
  {Koga}}, \bibinfo {author} {\bibfnamefont {N.}~\bibnamefont {Kawakami}},
  \bibinfo {author} {\bibfnamefont {T.~M.}\ \bibnamefont {Rice}}, \ and\
  \bibinfo {author} {\bibfnamefont {M.}~\bibnamefont {Sigrist}},\ }\href
  {\doibase 10.1103/PhysRevLett.92.216402} {\bibfield  {journal} {\bibinfo
  {journal} {Phys. Rev. Lett.}\ }\textbf {\bibinfo {volume} {92}},\ \bibinfo
  {pages} {216402} (\bibinfo {year} {2004})}\BibitemShut {NoStop}%
\bibitem [{\citenamefont {Huang}\ \emph {et~al.}(2012)\citenamefont {Huang},
  \citenamefont {Wang},\ and\ \citenamefont {Dai}}]{PhysRevB.85.245110}%
  \BibitemOpen
  \bibfield  {author} {\bibinfo {author} {\bibfnamefont {L.}~\bibnamefont
  {Huang}}, \bibinfo {author} {\bibfnamefont {Y.}~\bibnamefont {Wang}}, \ and\
  \bibinfo {author} {\bibfnamefont {X.}~\bibnamefont {Dai}},\ }\href {\doibase
  10.1103/PhysRevB.85.245110} {\bibfield  {journal} {\bibinfo  {journal} {Phys.
  Rev. B}\ }\textbf {\bibinfo {volume} {85}},\ \bibinfo {pages} {245110}
  (\bibinfo {year} {2012})}\BibitemShut {NoStop}%
\bibitem [{\citenamefont {Ylvisaker}\ \emph {et~al.}(2009)\citenamefont
  {Ylvisaker}, \citenamefont {Kune\ifmmode~\check{s}\else \v{s}\fi{}},
  \citenamefont {McMahan},\ and\ \citenamefont
  {Pickett}}]{PhysRevLett.102.246401}%
  \BibitemOpen
  \bibfield  {author} {\bibinfo {author} {\bibfnamefont {E.~R.}\ \bibnamefont
  {Ylvisaker}}, \bibinfo {author} {\bibfnamefont {J.}~\bibnamefont
  {Kune\ifmmode~\check{s}\else \v{s}\fi{}}}, \bibinfo {author} {\bibfnamefont
  {A.~K.}\ \bibnamefont {McMahan}}, \ and\ \bibinfo {author} {\bibfnamefont
  {W.~E.}\ \bibnamefont {Pickett}},\ }\href {\doibase
  10.1103/PhysRevLett.102.246401} {\bibfield  {journal} {\bibinfo  {journal}
  {Phys. Rev. Lett.}\ }\textbf {\bibinfo {volume} {102}},\ \bibinfo {pages}
  {246401} (\bibinfo {year} {2009})}\BibitemShut {NoStop}%
\bibitem [{\citenamefont {Werner}\ and\ \citenamefont
  {Millis}(2007)}]{PhysRevLett.99.126405}%
  \BibitemOpen
  \bibfield  {author} {\bibinfo {author} {\bibfnamefont {P.}~\bibnamefont
  {Werner}}\ and\ \bibinfo {author} {\bibfnamefont {A.~J.}\ \bibnamefont
  {Millis}},\ }\href {\doibase 10.1103/PhysRevLett.99.126405} {\bibfield
  {journal} {\bibinfo  {journal} {Phys. Rev. Lett.}\ }\textbf {\bibinfo
  {volume} {99}},\ \bibinfo {pages} {126405} (\bibinfo {year}
  {2007})}\BibitemShut {NoStop}%
\bibitem [{\citenamefont {Shim}\ \emph {et~al.}(2009)\citenamefont {Shim},
  \citenamefont {Haule},\ and\ \citenamefont {Kotliar}}]{shim_epl09}%
  \BibitemOpen
  \bibfield  {author} {\bibinfo {author} {\bibfnamefont {J.~H.}\ \bibnamefont
  {Shim}}, \bibinfo {author} {\bibfnamefont {K.}~\bibnamefont {Haule}}, \ and\
  \bibinfo {author} {\bibfnamefont {G.}~\bibnamefont {Kotliar}},\ }\href
  {http://stacks.iop.org/0295-5075/85/i=1/a=17007} {\bibfield  {journal}
  {\bibinfo  {journal} {EPL}\ }\textbf {\bibinfo {volume} {85}},\ \bibinfo
  {pages} {17007} (\bibinfo {year} {2009})}\BibitemShut {NoStop}%
\bibitem [{occ()}]{occupancy}%
  \BibitemOpen
  \href@noop {} {}\bibinfo {note} {Here, the $n_{5/2}$ and $n_{7/2}$ are
  evaluated using Matsubara Green's function $G(i\omega_n)$. The sum ($n_{5f} =
  n_{5/2} + n_{7/2}$) is somewhat smaller than that calculated from
  $\sum_{\Gamma}p_{\Gamma}N_{\Gamma}$.}\BibitemShut {Stop}%
\end{thebibliography}%

\end{document}